# Holography of Wi-Fi radiation

Philipp M. Holl (corr. author), Philipp.Holl@wsi.tum.de

Friedemann Reinhard, Friedemann.Reinhard@wsi.tum.de

Technische Universität München, Walter Schottky Institut and Physik Department, Am Coulombwall 4, 85748 Garching, Germany

Slide show showcasing the described method available at https://youtu.be/vle2ssAWiFg

**Abstract**

Wireless data transmission systems such as Wi-Fi or Bluetooth emit coherent light - electromagnetic waves with precisely known amplitude and phase. Propagating in space, this radiation forms a hologram – a two-dimensional wavefront encoding a three-dimensional view of all objects traversed by the light beam.

Here we demonstrate a scheme to record this hologram in a phase-coherent fashion across a meter-sized imaging region. We recover three-dimensional views of objects and emitters by feeding the resulting data into digital reconstruction algorithms. Employing a digital implementation of dark field propagation to suppress multipath reflection we significantly enhance the quality of the resulting images. We numerically simulate the hologram of a 10m-sized building, finding that both localization of emitters and 3D tomography of absorptive objects could be feasible by this technique.

## Introduction

Holography – three-dimensional imaging by phase-coherent recording of a two-dimensional wavefront – is one of the most intriguing concepts of 20th century physics[1]. While most practical implementations have employed laser light, the concept itself is applicable to any kind of coherent radiation and has actually been invented to improve electron microscopy. Other demonstrations have since been performed with sound waves[2], x-rays[3], gamma rays[4], neutrons[5] and cold atoms[6].

It is an interesting question whether the omnipresent stray radiation of wireless devices forms holograms that encode three-dimensional views of the device and its surrounding. So far, holography of microwave radiation with similar (GHz) frequency has been demonstrated for the localization of radio-frequency emitters in a two-dimensional plane[7,8] and near-field imaging with custom-built emitters[9,10]. However, efforts to obtain images from the stray radiation of unmodified wireless devices have remained limited to one-dimensional ranging[11].

Beyond its fundamental interest, indoor imaging by arbitrary wireless signals appears attractive for a variety of applications. These range from localization of radio-frequency tags in internet-of-things settings[7,8,12–14] over 3D motion capture for gaming[15–17] to through-wall-imaging of moving targets for security enforcement[11,18–21]. Unfortunately, it is complicated by one major challenge: multipath reflections – radiation scattered from walls and other surrounding objects outside the viewing area - blur radar signals and their echoes in indoor environments[14]. Existing approaches deal with this problem by producing or receiving short-pulse ultra-wideband signals for time-domain ranging[22–24] or directed radiation for beam scanning and cancellation of multipath interference[17,25,26]. In doing so they use only selected parts of the coherent radiation. To the best of our knowledge, imaging based on phase-coherent holographic recording of the full radiation field has remained elusive.

Here we demonstrate a holographic scheme to acquire three-dimensional images of building interiors from the radiation of an unmodified commercial narrowband Wi-Fi router. Our method does not require any prior knowledge of the emitted radiation and works with any type of signal, including encrypted communication.

## Setup and Data Acquisition

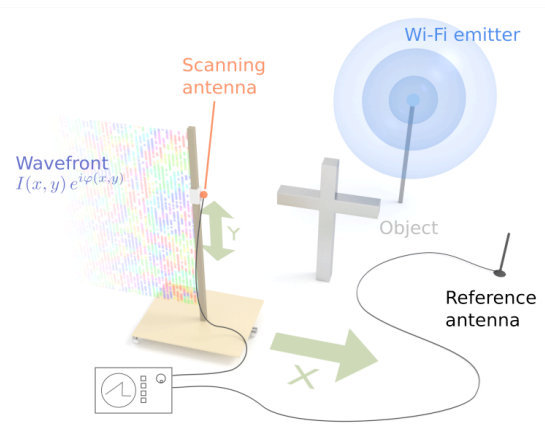

Figure 1: Experimental setup. Stray radiation from a commercial router is employed to image meter-sized objects (gray cross) by digital holography. We record a hologram of Wi-Fi radiation by a synthetic aperture approach. A Wi-Fi antenna (scanning antenna) is moved across a 3 by 2 meter plane, point-wise registering Wi-Fi signals. The signal phase is recovered by a homodyne

Our approach is presented in Figure 1. Additional details can be found in the supplementary material. We regard Wi-Fi radiation as coherent electromagnetic radiation – electromagnetic waves with well-defined amplitude $I$ and phase $\phi$. In this picture, a two-dimensional wavefront $I(x,y)e^{i\phi(x,y)}$ at any plane in space represents a hologram[1]. It encodes a three-dimensional view of all objects traversed by the light beam, which can be recovered by digital reconstruction[27]. We record such holograms of Wi-Fi radiation by scanning an antenna across a meter-sized two-dimensional plane ("scanning antenna" in Fig. 1) pointwise registering received Wi-Fi signals $I(t)$ in the time domain by an oscilloscope (Rigol DS4034 + analog demodulation, 350MHz bandwidth, see SI for details and Fig 2a for data).

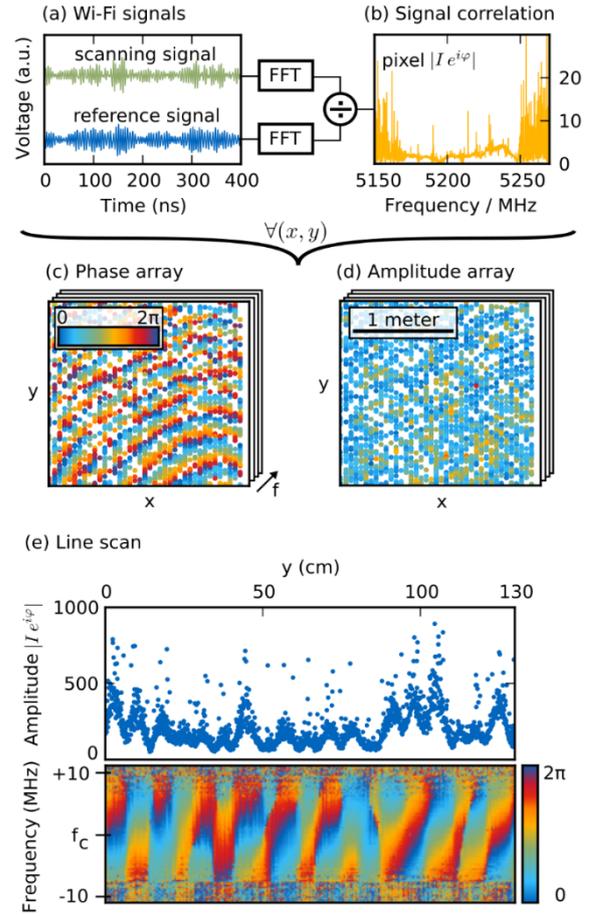

*Figure 2: Homodyne scheme for phase recovery. (a) The time-domain signals of the scanning and stationary antenna are Fourier-transformed and (b) normalized to each other to obtain amplitude attenuation I and the phase delay Φ imparted by the propagation for every frequency f within the Wi-Fi bandwidth. (c,d) Holograms are obtained by doing this for every pixel. (e) Multipath reflections in the building lead to a strong modulation with VSWR=9.7 and correlated artefacts in the phase. $f_c \approx 2.4$ GHz denotes the carrier frequency. (a-d) have been recorded using a 5GHz Wi-Fi emitter,(e) using a 2.4GHz emitter.*

We obtain the complex amplitude $I \cdot e^{i\Phi}|_{x,y}$ by a homodyne scheme, which is presented in Fig. 2a+b. In Fourier space, the propagation of the emitted complex signal $\widehat{I_{em}(f)}$ to the point $x,y$ is described as $U(x,y,f)\widehat{I_{em}(f)}$ where $U(x,y,f)$ denotes the propagation operator which includes amplitude attenuation and phase delay. As the emitter is essentially a black box, $\widehat{I_{em}(f)}$ is a priori unknown. To obtain the complex field, we therefore normalize the data using the signal received by a stationary reference antenna at position $x_r, y_r$ (Fig. 1), defining

$$I(f) \cdot e^{i\Phi(f)}|_{x,y} = \widehat{I(f)}|_{x,y}$$

$$\overset{\text{def}}{=} \frac{U(x,y,f)\,\widehat{I_{em}(f)}}{U(x_r,y_r,f)\,\widehat{I_{em}(f)}}$$

$$= \frac{\mathcal{F}\{I(t)|_{x,y}\}}{\mathcal{F}\{I(t)|_{x_r,y_r}\}} \quad (1)$$

Here, $\mathcal{F}$ denotes the temporal Fourier transform.

This scheme recovers amplitude $I$ and phase $\Phi$ of the propagated wavefront up to an irrelevant constant factor $U(x_r, y_r, f)^{-1}$ for all frequencies $f$ within the Wi-Fi channel employed (Figure 2b). Repeating this analysis for every pixel $(x, y)$ yields a hologram as displayed in Fig. 2 c+d, a 2D map of both phase and amplitude for each frequency $f$. All experimental datasets in this work have been recorded in a closed room with other objects and metallic walls and are therefore strongly affected by standing waves from multipath reflection. We benchmark their strength from a single line of a 2D scan (Fig. 2e). The amplitude modulation has a standing wave ratio of SWR=9.7. It creates artefacts in the recovered phases, which correlate with peaks and valleys of the amplitude.

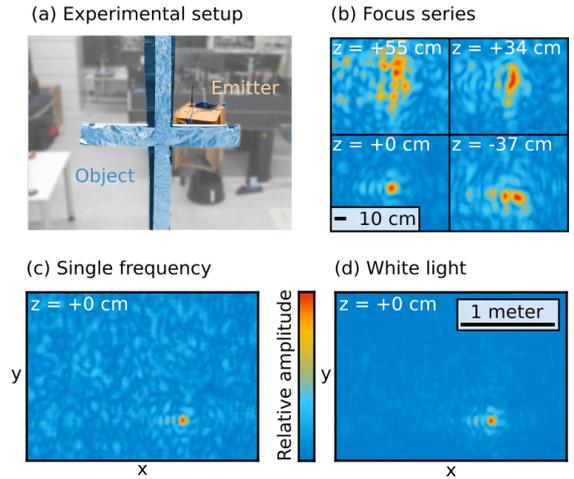

Figure 3: Recovery of 3D holographic views by numerical back propagation of the hologram. (a) A commercial Router is used to illuminate a cross-shaped phantom object. (b) Back-propagation into the emitter plane reliably reveals the router as a single bright spot. (c) Multipath reflections create a speckle pattern that (d) can be cancelled by incoherently averaging over holograms obtained at different transmission frequencies.

We now turn to the three-dimensional reconstruction of images from this holographic data. All following analysis is based on a dataset, which has been recorded in the setting of Figure 3a. We capture a metallic cross-shaped phantom object. The illumination source is a commercial 5GHz Wi-Fi emitter (TP-LINK Archer C20, 802.11ac), placed at a distance of $z_e = 230$cm from the recorded plane, 90cm behind the object plane $z_o$. Image reconstruction is performed by numerical backpropagation. This scheme recovers the light field at an arbitrary depth $z$ by propagating the recorded holographic wavefront from $z_0$ in space according to the angular-spectrum relation[28,29]

$$I(x,y,z)$$
$$= \mathcal{F}^{-1}\left\{\exp\left[\mp i\,\frac{2\pi(z-z_0)}{\lambda}\sqrt{1-\lambda^2 f_x^2 - \lambda^2 f_y^2}\right]\right.$$
$$\left.\times \mathcal{F}\big(I(x,y,z_0)\big)\right\} \quad (2)$$

Here, $\mathcal{F}, \mathcal{F}^{-1}$ denote the spatial Fourier transform of the light field and its inverse and $f_x, f_y$ the transform's spatial frequencies. The sign in the exponential function is chosen negative by default and positive for $1 - \lambda^2 f_x^2 - \lambda^2 f_y^2 < 0$ to prevent exponential blow-up of evanescent waves. Backpropagation for a range of $z$ yields a 3D picture of all emitters and objects traversed by the emitted beams.

**Analysis and Results**

The most prominent feature in the recovered 3D view is the emitter (Figure 3b-d), which appears as a bright spot at the expected distance ($z_e = 230$cm). Its extension ($\sigma_x = 3$cm) is close to the diffraction limit set by the size of the recording plane and comparable to the Wi-Fi wavelength, allowing for cm-scale localization of its position in 3D space. Remarkably, this high accuracy appears unaffected from the presence of multipath reflections. We explain this effect by the fact that the freely propagating part of radiation converges to a single point at $z_e$, while all scattered radiation loses its spatial information, resulting in an unspecific extended interference pattern of speckles at $z_e$. This view is supported by the observation that the emitter focus (bright spot in Fig 3b-d) contains only 15% of the received intensity.

We further reduce the intensity of speckles by incoherent white light holography[30] (Figure 3d), summing the back-propagated light fields of every frequency $f$ as $I(x,y) = \sum_f |I(x,y,f)|^2$. This procedure suppresses speckle interference by exploiting the fact that Wi-Fi radiation actually is white light. This is due to its modulation with bit-patterned data, which reduces coherence length to a value set by the inverse signal bandwidth $\Delta f$, in our case ($\Delta f \approx 70$MHz) to a numerical value of $\Delta x = \frac{c}{\Delta f} = 4.3m$. Our scheme can recover a white light image from reconstructions at each frequency, similar to filtering techniques in digital holography[31]. In contrast to some of these competing techniques[32], it does not degrade spatial resolution, since it preserves phase information for each frequency component upon acquisition and performs incoherent averaging only on the level of reconstructed images.

Objects in space are expected to appear as a shadow in the reconstruction of the emitted light cone. Indeed, a white light reconstruction of the object plane $z_O$ reveals a shadow that correctly reproduces both the shape and dimensions of the cross-shaped phantom structure (Figure 4a+b). Speckles from multipath reflections are clearly visible despite the use of white-light suppression. Their detrimental effect on the image is stronger than in the case of the emitter plane (Figure 3), since the signal intensity is now spread across a larger cross section.

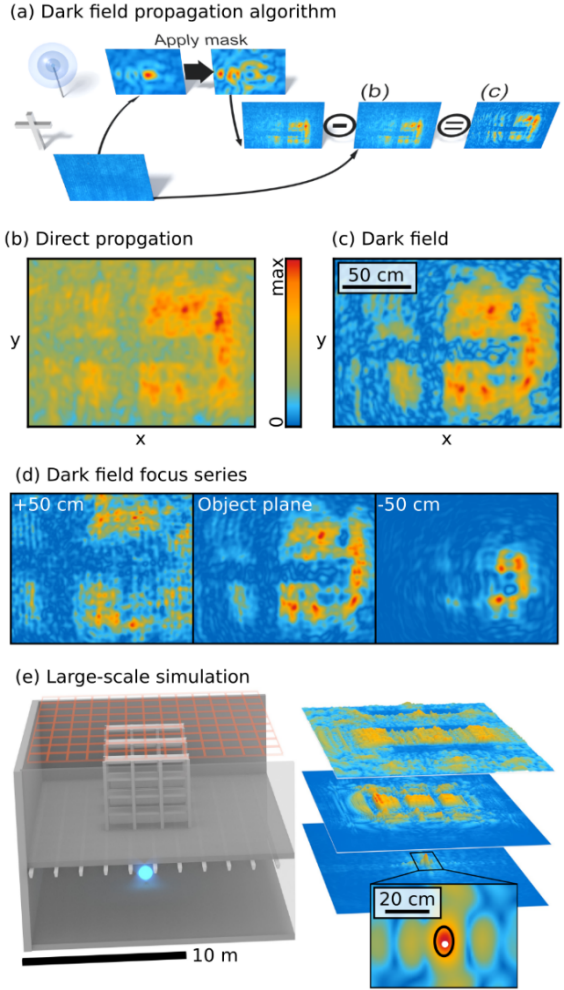

*Figure 4: Reconstruction of objects. (a) Sketch of dark field algorithm to enhance shadow contrast: a direct back-propagation into the object plane (b) is subtracted from the forward scattered radiation at this point, obtained by masking the emitter in its plane and forward-propagating the result. (c) Resulting dark field image, contrast of objects is visibly enhanced. (d) Objects appear blurred in defocussed planes, underlining the three-dimensional information contained in the hologram. (e) Simulation model of a full building on the left and reconstructed planes using our algorithm on the right. The black ring in the emitter plane shows the FWHM in intensity and the white dot the true emitter location.*

We can further increase imaging contrast by a digital implementation of dark field propagation. This technique is commonly used in microscopy to reveal weakly absorbing objects. It reveals the scattered light field of an object by blocking out the illumination source by a suitable spatial filter.

Our digital version of the dark field method eliminates the illumination source by masking the Wi-Fi source in the emitter plane, as shown in figure 4a. The remaining field is then propagated to the object plane and subtracted from the directly reconstructed image to obtain

$$|I_{Darkfield}(x,y,z)|^2 = |I_{Masked}(x,y,z)|^2 - |I_{Unmasked}(x,y,z)|^2 \quad (3)$$

By subtracting the two fields, one filters out the missing direct illumination which constitutes the dark field. Indeed, dark field propagation reveals structures with a higher contrast, as can be seen from a comparison to direct reconstruction (figure 4b+c).

It is important to note that, with this method, points outside the dark field cone are completely unphysical which can be seen in figure 4b-d. As the emitter cannot be traced back to a single point but instead is spread out over a few centimeters, we use a Gaussian shaped mask to block out its contribution. This comes at the disadvantage that a larger mask will produce a narrower dark field beam. To overcome that problem, we only block out a part of the direct light. This results in a wider cone and produces a field that is approximately scaled by a constant.

Figure 4 d) shows reconstructions for different depths. The image of the structure appears blurred in out-of-focus planes, underlining the three-dimensional information contained in the hologram.

As a final outlook, we investigate the feasibility of a realistic large-scale reconstruction and tracking application (Fig. 4e). Therefore, we performed a finite difference time domain simulation of the radiation field of a pointlike dipole emitter (frequency $f = 2.4$GHz, bandwidth $\Delta f = 20$MHz) on the floor of a 20x17x12m$^3$ storage building. We record a hologram across a two-dimensional plane in the ceiling (orange grid), as it could be implemented in practice by a two-dimensional array of stationary antennae. As a model for objects we add an intermediate floor at a height of 6m, supported by perfectly reflective metal bars, which supports a set of perfectly reflective metallic shelves. The three-dimensional reconstruction of the hologram clearly shows sharp outlines of the metallic shelves and bars in their respective reconstructed planes. The emitter location can be inferred with an accuracy of only 3cm, determined by the FWHM of intensity in its pointlike image, despite the presence of multipath reflections from the surrounding structures.

## Conclusions

Several future improvements of the technique appear attractive. First, a stationary two-dimensional array of antennae could replace our slow synthetic aperture approach of mapping a wavefront by pointwise scanning. Such a device could provide pictures several orders of magnitude faster. Specifically, video rate 3D imaging appears feasible, since a 1GS/s acquisition device could acquire a 100ns scale snapshot of signals for one million pixels at a rate of 10 frames/s. Second, future wireless signals will significantly improve image quality. Their higher bandwidth will reduce speckle contamination, while their higher frequencies will increase spatial resolution, to the 5mm scale for 60 GHz transmission currently under discussion (IEEE 802.11ad).

The implications of this result are manifold. It opens a path to three-dimensional imaging and localization in presence of multipath reflections, which does not require ultra-wideband signals or directed emitters. Its power in 3D localization of sources could find applications in indoor navigation, where it could track radio-frequency labels even in cluttered indoor settings, potentially at video rate acquisition speeds. It equally raises concerns about the privacy of wireless communication: even encrypted communication transmits a three-dimensional picture of its surrounding to the outer world, which can be recovered by suitable strategies.

## Acknowledgements

This work has been supported by the Deutsche Forschungsgemeinschaft (Emmy Noether grant RE 3606/1-1).

# Supplemental Material

**Data acquisition and electronics**

All electronic components are listed in table 1. Devices used for mixing and amplifying are from Mini Circuits UK. As the oscilloscopes used in this experiment cannot resolve the high-frequency Wi-Fi radiation directly, we down-convert the signals using a high frequency oscillator as shown in figure 5.

Depending on the antenna type and cable length, the incoming signals may be too weak to obtain a reasonable voltage resolution. We therefore amplify the high-frequency signals before they are sent to the mixing stage. The signal distortions introduced by this setup are negligibly small and do not impact later analysis.

The mixing stage consists of a high-frequency local oscillator set to approximately 2.4 and 5 GHz, respectively. Its output is split and then fed into two equal mixers, one for each antenna. We have chosen the cables in this step to be identical for both paths. The arbitrary relative phase introduced this way is irrelevant for the later analysis.

Data recorded by the oscilloscope are then transferred to the on-board PC via USB. As the USB interface is not fast enough to deliver the data in real time, a trigger is set on the reference signal whose amplitude is constant during the scanning process. The trigger setup limits the data acquisition rate to about two points per second on average.

**White light reconstruction**

We suppress multipath interference by white light holography as presented in Fig. 6. In this scheme, the final reconstructed image (Fig 6 "combined") is an incoherent sum of reconstructions obtained at different frequencies within the Wi-Fi transmission band (illustrated in the remaining subplots in Figure 6). We can generate separate holograms for arbitrary frequency windows within the transmission band, since our acquisition scheme (Fig. 2) records a frequency-resolved three-dimensional (x,y,f) dataset.

Multipath reflection adds unpredictable noise to the recording, which transforms to speckle patterns in the frequency-windowed holograms. Crucially, these patterns vary randomly between different frequency bands and combine to a homogeneous background in the combined image. Multipath suppression and image clarity improve for increasing bandwidth. This is similar to time-domain ranging, where shorter pulses achieve the same goal [1].

We note that our approach does not suffer from artefacts frequently observed in other white light holography schemes, such as a reduced depth of field [2]. The key difference is that our scheme preserves phase information for each frequency and therefore enables fully coherent back-propagation, while other techniques perform incoherent averaging already upon recording. In this way, our work is conceptually more similar to holographic imaging with filtered white light [3].

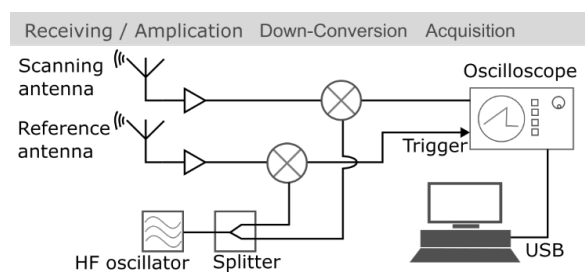

*Figure 5: Circuit diagram for data acquisition.*

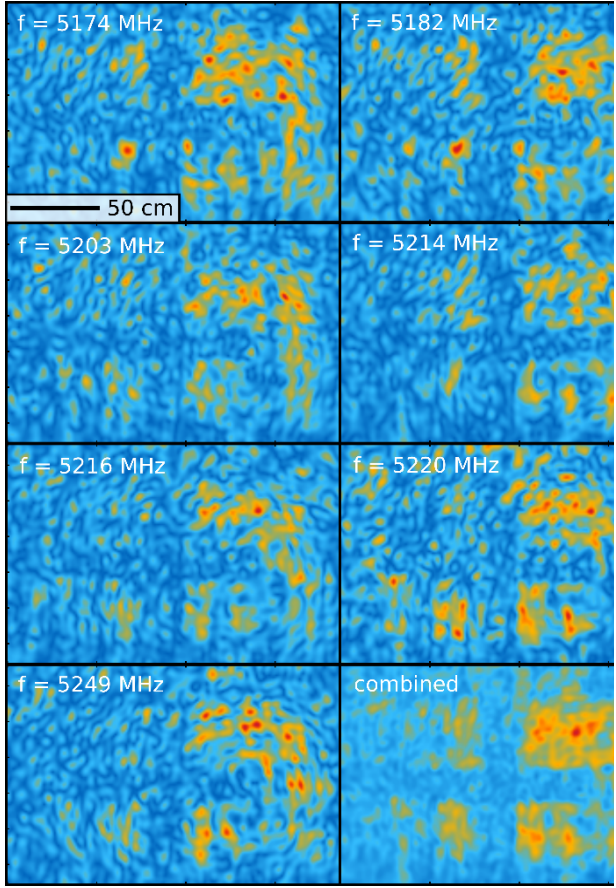

*Figure 6: White-light holographic reconstruction of the object plane using different frequencies. Each image shows the back-propagation of a frequency band with a width of 1.6 MHz. The last image is the RMS mixture of 50 individually reconstructed holograms ranging from 5170 MHz to 5250 MHz.*

**FDTD simulation**

To validate our method and show the reconstruction limits, we simulate several scenarios, both with 5 GHz and 2.4 GHz Wi-Fi, using the commercially available Lumerical finite differences time domain (FDTD) toolkit. The storage hall is the most complex model simulated here.

The limiting factor for our type of FDTD simulation is the system memory requirement. We therefore simulate 2.4 GHz Wi-Fi instead of the higher-bandwidth 5 GHz which is used for the experimental results. The meshing size of 1.6 to 2.4 cm (homogeneous in space but different per axis) is set as large as possible while keeping the necessary precision for reconstruction, making sure the meshing size stay under 20% of the shortest occurring wavelength. With these settings, the simulation requires just under 80 GB of memory.

For the source, we use a dipole emitter producing a single wave packet with a bandwidth of 22 MHz centered around the carrier frequency in accordance with the 802.11 standard. For the different source positions, also the Wi-Fi channels are varied. The simulation time of 300 ns ensures that any non-negligible multipath reflections are received by the antenna array.

The antenna array, located in the ceiling of the building, is realized as a field time monitor with a resolution of 4.0 cm x 7.2 cm and a sampling rate of 5.7 GHz.

Figure 7a) shows a series of reconstructed planes with the emitter located approximately in the center of the room. The cross sections clearly reveal the shapes of both the metallic shelves and the metallic bars underneath.

The localization of source can be broken into finding the correct distance from the recording plane and a 2D localization in the respective plane. The 2D localization is very simple as the intensity distribution approximately forms a Gaussian near its maximum. The standard deviation of this Gaussian is of the order of a few centimeters at a depth of z=11 meters.

Finding the correct plane, however, is more difficult as the peak in intensity rapidly changes shape over the range of a few decimeters. The evolution of the peak broadness over a range of depths close to the source position is plotted in figure 7b). The variances in X and Y direction both have a minimum corresponding to the most likely z position of the emitter. With their minima at z=10.85 and z=11.2 meters, they are a few decimeters off the true source position at a depth of 10.6 meters. One possible reason for this might be the 38 cm thick ceiling, through which the waves pass before reaching the antenna array.

(a) Reconstructed planes at different depths

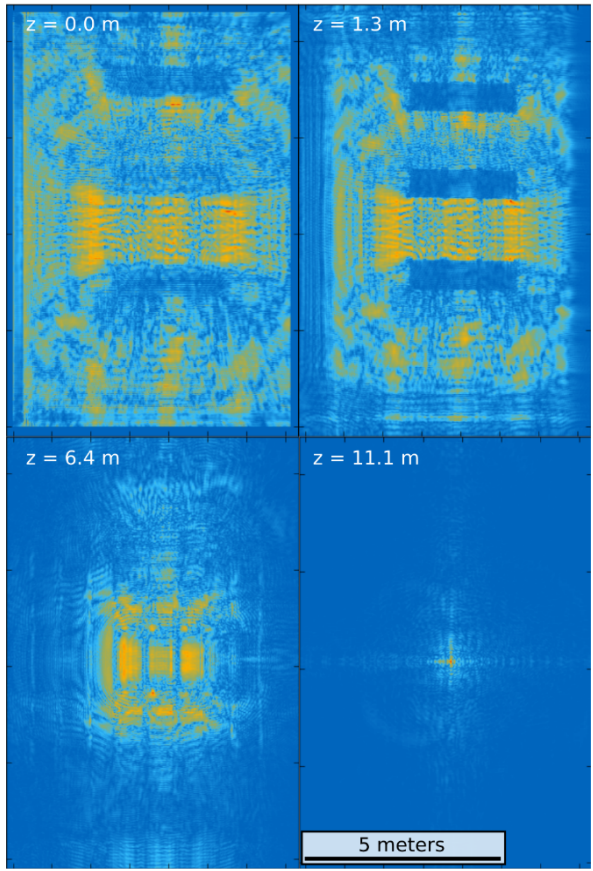

(b) Depth localization of the emitter

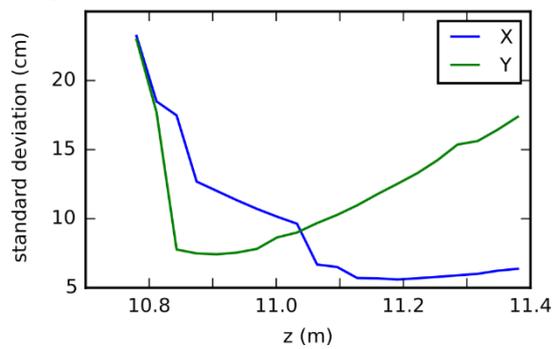

*Figure 7: Reconstruction of simulated storage hall.*

| DEVICE TYPE | 2.4 GHZ | 5 GHZ |
| --- | --- | --- |
| EMITTER | Google Nexus S in hotspot mode, 2.4 GHz | TP-LINK Archer C20, 802.11ac, 5 GHz |
| REFERENCE ANTENNA | INTELLINET NETWORK SOLUTIONS Indoor Omni-Directional Antenna, 2.4 GHz, 5 dBi, Part number 502290 | Delock 88899, 7 dBi 2.4 GHz, 5 GHz |
| SCANNING ANTENNA | Custom design | Laird Technologies IAS MAF94264 |
| AMPLIFIER | Mini-circuits ZX60-43-S+ | Mini-circuits ZX60-8008E-S+ |
| MIXER | Mini-circuits ZX05-C42-S+ | Mini-circuits ZX05-14-S+ |
| HF OSCILLATOR | Mini-circuits ZX95-2500A-S+ | Lab Brick Signal generator LMS-103 Opt-004, 5-10 GHz |
| SPLITTER | Mini-circuits ZX10-2-442-S+ | Mini-circuits ZX10-2-71-S+ |
| BAND PASS FILTER | Mini-circuits VBF-2435+ | Mini-circuits VBF-2435+ |
| OSCILLOSCOPE | Rigol DS4034, 350 MHz bandwidth, 4 GSa/s<br>Rigol DS1052E, 50 MHz bandwidth, 500 MSa/s | Rigol DS4034, 350 MHz bandwidth, 4 GSa/s |

Table 1: Devices used in 2.4 GHz and 5 GHz setup.